\documentclass[sigconf]{acmart}
\usepackage{adjustbox}

\AtBeginDocument{%
  \providecommand\BibTeX{{%
    \normalfont B\kern-0.5em{\scshape i\kern-0.25em b}\kern-0.8em\TeX}}}

\setcopyright{acmcopyright}
\copyrightyear{2021}
\acmYear{2021}
\acmConference[KCAP 21]{KCAP 2021}

\begin{document}

\title{Local Explanations for Clinical Search Engine results}

\author{Edeline Contempré}
\email{edeline_contempre@yahoo.com}
\orcid{0000-0002-1767-121X}
%\email{webmaster@marysville-ohio.com}
\affiliation{%
  \institution{Vrije Universiteit Amsterdam}
  \streetaddress{De Boelelaan 1105}
  \city{Amsterdam}
  \country{the Netherlands}
}

\author{Zoltán Szlávik}
\email{z.szlavik@vu.nl}
\orcid{0000-0002-2781-3795}
\affiliation{%
  \institution{Vrije Universiteit Amsterdam}
  \streetaddress{De Boelelaan 1105}
  \city{Amsterdam}
  \country{the Netherlands}}

\author{Majid Mohammadi}
\email{m.mohammadi@vu.nl}
\orcid{**}
\affiliation{%
  \institution{Vrije Universiteit Amsterdam}
  \streetaddress{De Boelelaan 1105}
  \city{Amsterdam}
  \country{the Netherlands}}

\author{Erick Velazquez}
\email{erick.velazquez@mytomorrows.com}
\affiliation{%
 \institution{Vrije Universiteit Amsterdam}
 \streetaddress{De Boelelaan 1105}
 \city{Amsterdam}
 %\state{Arunachal Pradesh}
 \country{the Netherlands}}

\author{Annette ten Teije}
\email{annette.ten.teije@vu.nl}
\affiliation{%
  \institution{Vrije Universiteit Amsterdam}
  \streetaddress{De Boelelaan 1105}
  \city{Amsterdam}
  %\state{Texas}
  \country{the Netherlands}}
%\email{cpalmer@prl.com}

\author{Ilaria Tiddi}
\email{i.tiddi@vu.nl}
\affiliation{%
  \institution{Vrije Universiteit Amsterdam}
  \streetaddress{De Boelelaan 1105}
  \city{Amsterdam}
  \country{the Netherlands}}
%\email{jsmith@affiliation.org}

\renewcommand{\shortauthors}{E. Contempré, et al.}

\begin{abstract}
Health care professionals rely on treatment search engines to efficiently find adequate clinical trials and early access programs for their patients. However, doctors lose trust in the system if its underlying processes are unclear and unexplained. In this paper, a model-agnostic explainable method is developed to provide users with further information regarding the reasons why a clinical trial is retrieved in response to a query. To accomplish this, the engine generates features from clinical trials using by using a knowledge graph, clinical trial data and additional medical resources. and a crowd-sourcing methodology is used to determine their importance. Grounded on the proposed methodology, the rationale behind retrieving the clinical trials is explained in layman's terms so that healthcare processionals can effortlessly perceive them. In addition, we compute an explainability score for each of the retrieved items, according to which the items can be ranked. The experiments validated by medical professionals suggest that the proposed methodology induces trust in targeted as well as in non-targeted users, and provide them with reliable explanations and ranking of retrieved items.
\end{abstract}

\keywords{Explainability, Search, Health care, Crowdsourcing, Treatment search}
\maketitle

%%%%%%%%%%%%%%%%%%%%%%%%%%%%%%%%%%%%%%%%%%%%%%%%%%%%%%%%%%%
%%%% Introduction
%%%%%%%%%%%%%%%%%%%%%%%%%%%%%%%%%%%%%%%%%%%%%%%%%%%%%%%%%%%
%\textbf{My questions:}
%\begin{itemize}
%    \item What do you think about the title? should be "Search Engines for Clinical Trials" or "Search Engines for Health Systems"?
    
%    \item I think we need to have one/two paragraphs at the beginning of the introduction just to lay the ground for the further discussion! Right now we rather start the problems rather abruptly! 
    
%    \item Except for user-friendly explanations, what is the issues related to LIME or SHAP for our problem?
%\end{itemize}

\section{Introduction} \label{introduction}

When healthcare professionals (HCPs) use a treatment search engine to find treatment options for their patients, they need to gain a certain trust in the system. While accuracy, performance, and the design are essential for accomplishing such trust, we may need to do more to reach a threshold where HCPs trust the system enough to want to use it in critical scenarios, e.g., when a patient’s life may be at risk.

A search engine would typically provide a ranked list of the related items with regards to a certain query. However, lack of explanations could lead to a lack of trust from users as they would not understand the underlying logic of retrieving an item in response to a query. In the medical domain, where the pressure to make no mistakes is high,incorrectly attributing the cause of a mistake could be fatal.  As a result, without the ability to interpret the model, HCPs’ trust in the model decreases and will, ultimately, not use the model’s outputs \cite{pu2006trust}. In addition, due to compliance regulations, most search engines in the medical domain provide unordered lists of related items in response to a query, making it difficult for users to look into or distinguish the most relevant items for their needs. As an instance, treatment search engines such as clinicaltrials.gov do not offer relevance-based ranking of documents, partly because ranking in-criteria treatment options may be suspect to favoritism, which is highly prohibited for clinical trials \cite{Clinical58:online}. In addition, for an efficient and thorough search, a status/date/title-based ordering may not always be the most practical for end users as they tend to scatter similar results from each other.

A potential solution to provide explanations is to use current explainable methods. A class of such methods would provide global explanations only so that we can evaluate the overall search engine as a whole, but cannot provide proper explanations in response to each individual query \cite{dam2018explainable}. Another option is to use other methods for local explainability such as LIME \cite{das2020opportunities} and SHAP \cite{lundberg2017unified} as these explain how or why a specific result is provided. However, these methods are designed for machine learning problems, such as classification and regression, but not for explaining search engine results. When local explanations were used to explain search results, these were either based on one feature from the documents like word prominence \cite{verma2019lirme}, or not applicable to the clinical trials since they are based on user reviews for each document \cite{catherine2017explainable} (which are unavailable for clinical trials). 

Another major challenge is to provide HCPs with user-friendly, reliable, and easy-to-understand explanations. A major drawback of current explainability techniques, such as LIME \cite{das2020opportunities}, is that these techniques tend to focus on aiding users with technical backgrounds to interpret the system. HCPs are not universally expected to understand the detailed workings of a complex retrieval model. Thus, HCPs require explanations that are high level and intuitive, while they do not necessarily need to reflect the exact inner workings of retrieval models. This opens up an opportunity in which explanation models can be built in a way that they are not sensitive to minor changes in the model of a search engine. For instance, with a treatment search engine, we may offer a local explanation such as “the queried disease is mentioned in the retrieved clinical trial’s title”, and as long as the retrieval model relies on this as a feature, we are using a generic explanation method, and we can develop the search engine and explanation module fairly separately in practice. 

In this paper, an explainability method is developed which provides tailored explanations to medical practitioners for retrieved items. To that end, meaningful features from clinical trials are extracted from different data sources, and preferences of different users are elicited by utilizing a crowdsourcing-based methodology. We then put forward a method to translate preferences into importance level of features. Based on features' importance levels, tailored explanations are acquired for each specific query, according to which we develop a sentence template in order to present them to users. In addition, we introduce explainability scores, according to which we order retrieved items. The results suggest that the use of local explainability on clinical search engines promote HCPs trust, search experience, and result ordering satisfaction. 

%The remaining of this article is structured as follows. Section \ref{sec:related works} presents the related work in the search engines in the medical domain. Section \ref{sec:method} is dedicated to the proposed method including feature engineering and importance identification, and explaining as well as ranking the retrieved items in response to a query. Section \ref{sec:experiments} contains the experiment regarding the proposed methodology, and the paper is finally concluded in Section \ref{sec:conclusion}.

%%%%%%%%%%%%%%%%%%%%%%%%%%%%%%%%%%%%%%%%%%%%%%%%%%%%%%%%%%%
%%%% Related Work
%%%%%%%%%%%%%%%%%%%%%%%%%%%%%%%%%%%%%%%%%%%%%%%%%%%%%%%%%%%
\section{Related Work}\label{sec:related works}
%% It is good to discuss what a clinical trial is! Later on we really need to use it!
%%%%%%%%%%%%%%%%%%%%%%%%%%%%%%%%%%%%%%%%%%%%%%%%%%%%%%%%%%%
%%%% Explainable Recommender System
%%%%%%%%%%%%%%%%%%%%%%%%%%%%%%%%%%%%%%%%%%%%%%%%%%%%%%%%%%%
%The growth of the number of publications in Pubmed about explainability is highlighted in figure \ref{articles} where we observe a steady growth of the use of XAI between 2017 and 2020. In 2017, 8 articles were written on XAI. In 2018, 12 articles were written on XAI. In 2019, 29 articles were written on XAI, twice as many compared to 2018. Lastly, 87 articles were written on XAI in 2020, nearly three times as much compared to 2019. Subsequently, research in explainability is increasing every year which stresses the relevance of the matter. 

% such as finance \cite{bracke2019machine, burgt2020explainable}, health care \cite{holzinger2019causability, london2019artificial, hossain2020explainable}, agriculture \cite{wolanin2020estimating} and more. 

The General Data Protection Rule (GDPR) requires (since 2016) all systems collecting data to be transparent on how these use data. Researchers have since then created surveys on explainability \cite{zhang2018explainable, dovsilovic2018explainable, adadi2018peeking}, defined different types of explainability \cite{sheh2018defining}, used explainability in a variety of sectors. However, up to date, there is no definition of explainability in the Oxford Dictionary\footnote{See https://www.oxfordlearnersdictionaries.com/spellcheck/english/?q=explainability (visited on 15/10/2020)} \cite{rosenfeld2019explainability}. Researchers argue about the definition of explainability, what shape explainability has, meaning it is a monolithic concept. 

To clarify our vision of explainability, we identified three main dimensions of explainability that can be observed throughout researchers' definitions: \textit{audience}, \textit{understanding}, and \textit{transparency}. \textit{Understanding} refers to the user's ability to understand the model's results. However, not all users can interpret all models using explainability as models can be domain specific. For example, users without knowledge in biology would struggle to understand highly biological terms generated by a model's explainability attempting to diagnose a certain type of lung cancer. Likewise, explainable AI (XAI) could use simple terms, leading to a lack of details for the doctor assessing the diagnosis. The user is therefore required to have a certain amount of knowledge to understand the explanation itself, making it crucial for developers using explainability to target their \textit{audience} \cite{rosenfeld2019explainability}. Lastly, an explainable method should increase the model's \textit{transparency} by making it more interpretable for its users, and not try to generate seemingly arbitrary explanations that do not fit with how the model works  \cite{dimanov2020you}.

A type of audience are, for example, health care professionals (HCPs) as they increasingly rely on artificial intelligence models to make concise decisions such as diagnosing or providing treatment options. A further increase of the use of AI occurs when the machine provides explanations as HCPs use these elaborations as a means to justify their decisions. For example, if a model suggests diagnosing a patient with lung cancer, the HCP would want a justification to this decision/suggestion. Other HCPs have reported a need to understand the cause of learned representations \cite{holzinger2019causability}. Moreover, HCPs reported it will not cause major complications if the model is not as accurate, but would always want to know its potential shortcomings. Therefore, local explanations benefit HCPs to construct their conclusions. 

Local explanations increase trust in HCPs. Although providing local explanations about features is a straightforward solutions, these do not aid all clinicians in all departments. HCPs working in the intensive care unit and in the emergency department reported that local explanations were not useful. The two departments have time as common denominator. HCPs in these departments often lack time due to patients with a short prognosis that only have a few hours to live if the HCP does not take action. Local explanations could, therefore, not be useful to users lacking time. 

Current state-of-the-art local explainability techniques do not use user-friendly explanations. These current local explainability techniques are either based on feature importance such as LIME \cite{das2020opportunities} and SHAP \cite{lundberg2017unified}, rule-based \cite{verma2019lirme}, saliency maps \cite{mundhenk2019efficient}, prototypes \cite{gee2019explaining} example based \cite{dave2020explainable}, or on counterfactual explanations \cite{dave2020explainable}. Up to date, feature importance \cite{zhang2019sigir} and rule-based techniques \cite{verma2019lirme} were used on search engines, but do not meet the criteria that these should be user friendly.

% \subsection{Overview on local explainability techniques}\label{background_explainability-stakeholders}\hfill\\

% Current state of the art local explainability techniques do not use user friendly explanations. These current local explainability techniques are either based on feature importance such as LIME \cite{das2020opportunities} and SHAP \cite{lundberg2017unified}, rule based \cite{verma2019lirme}, saliency maps \cite{mundhenk2019efficient}, prototypes \cite{gee2019explaining} or example based \cite{dave2020explainable}, or on counterfactual explanations \cite{dave2020explainable}. Up to date, feature importance \cite{zhang2019sigir} and rule based techniques \cite{verma2019lirme} were used on search engines, but do not meet the criteria that these should be user friendly. 

%\subsection{Local Interpretable Model agnostic Explanation (LIME)} \label{LIME}\hfill\\

%mention fidelity and the problems that have been encountered with it 

LIME is a type of local explainability method aiming to increase transparency for specific decisions given by an opaque model. It explains single result by letting users know why they are getting this specific result over another \cite{verma2019lirme}. Although LIME offers one way to solve the black-box problem, it has a few limitations. The first limitation of using LIME is that it is most commonly used for linear or classification models \cite{arrieta2020explainable}. This limits the degree to which the model can be meaningfully applied to, and restricts itself to non-user-friendly explanations. Consequently, this research does not use LIME methods, but developed a local explainability method to order and generate user friendly explanations. 

%, how the method is getting this specific result \cite{singh2019exs}, etc. Its input mainly restricts itself to sequence prediction, regression, and classification tasks \cite{singh2019exs}. For instance, LIME has been used to explain image classifications \cite{ribeiro2016should}, machine learning predictions and classification models \cite{robnik2008explaining}, or nonlinear classification algorithms \cite{baehrens2009explain}. In court, LIME has, for instance, been used to understand a model's decision on whether the defender was guilty or innocent. Adding LIME to the model showed that the model's decisions were highly influenced by race and gender. Hence, LIME helps detect bias in a model.

%Moreover, several articles called out the unfairness of LIME \cite{dimanov2020you, white2019measurable} as its feature selection is biased because it looks for the most important features. Research in \cite{dimanov2020you} targets low feature importance and demonstrate that a model significantly decreases relative importance, and that this decrease in relative importance is shown across all feature selection explainability model \cite{dimanov2020you}. In addition, research pointed out that LIME is unfaithful to its own model as some key features are most often not explainable \cite{white2019measurable}. 

%LIME approaches depend on the black box model itself (sequence prediction, regression, classification) ***. \cite{singh2019exs}

%\subsubsection{LIME limitations} \hfill\\

\section{Explainable Search Engine}\label{sec:method}

%Need to describe the crowdsourcing experiment in a fair amount of detail as the explainability scores depend on these. However, we need to keep in mind that we already wrote a paper with the findings of the crowdsourcing task.
%we retrieve the items, but we are not recommending it
%we have explanations, and they are further emphasized with ordering - consistency between explanations and ordering, where the ordering backs-up the experiment

This section presents the proposed model that provides explanations for its users, as well as how it orders a clinical search engine's results. This enables users to efficiently find potential relevant clinical trials while understanding the underlying processes of the model. The proposed method also generates local explainability scores for each clinical trial and uses these scores to order the search engine's results. Moreover, users are shown user-friendly explanations providing descriptions of the features are available in each clinical trial. The prerequisites of computing the explanations and ordering is the engineering of features from different data sources. 

\begin{figure}[H] 
    \centering
    \includegraphics[width=0.5\textwidth]{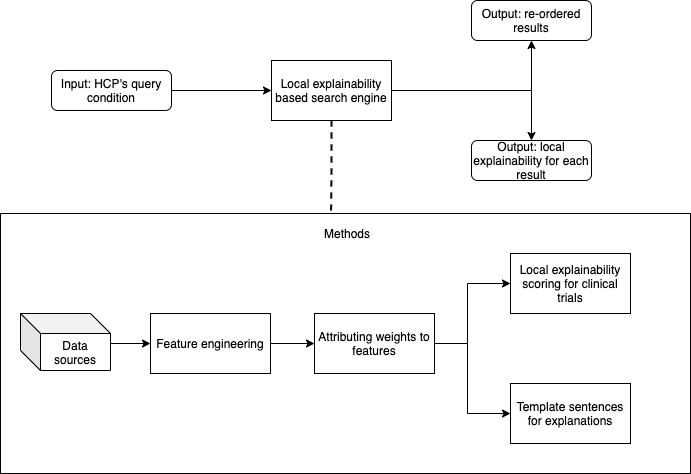}
    \caption{Overview of the methods' pipeline. Resources include the knowledge graph, data from UMLS, clinical trials from CT.gov, and data from Pubmed.} 
    \label{fig:pipeline}
\end{figure}

Figure \ref{fig:pipeline} shows the pipeline of steps conducted for the proposed methodology. The search engine takes as input the user's query, and returns an output with explainability-based ordered results with explanations. Figure \ref{fig:pipeline} shows that the local explainability search engine combines resources with the HCP's query to engineer features. These features are, thereupon, attributed local explainability scores which are used to order the list of clinical trials. In addition, the engineered features' outputs fill template sentences. These explanations provide information to the user on \textit{how much of this clinical trial can the search engine explain}. In the following section, each module in Figure \ref{fig:pipeline} is discussed and explained in more detail.

%\textbf{The figure needs to be changed according to the story of the article!}

%%%%%%%%%%%%%%%%%%%%%%%%%%%%%%%%%%%%%%% Feature Engineering

\subsection{Feature Engineering}

%Write in more detail how we engineered the features (in comparison to the crowdsourcing paper)

Before engineering the features linked to clinical trials, we first present the data sources according to which features for each clinical trial are extracted. 

Table \ref{tab:methods_data_used} provides an overview of the data sources used in the proposed model. First, we used data from UMLS (Unified Medical Language System) \cite{umls}, which is an official medical database where all conditions, diseases, infections, and more, are associated to Concept Unique Identifiers (CUIs). Second, different clinical trial sources such as clinicaltrials.gov\footnote{https://www.clinicaltrials.gov/} are used as it is the biggest clinical trial repository. Third, the database comprises Pubmed publications, as it saves medical papers. Lastly, it comprises data from the company's conditions graph (knowledge graph) where parent-child relations between diseases are defined, terms are specified, as well as the clarification of terms and their synonyms.

\begin{table}
\centering

\caption{Description of the data sources for the feature engineering.}
\label{tab:methods_data_used}

\begin{tabular}{l|p{5cm}}
\textbf{Data source} & \textbf{Data} \\
\hline
\textbf{UMLS} & Concept Unique Identifiers, disease terms, and relations between these \\
\textbf{clinicaltrials.gov} & Clinical trials' detailed descriptions, summaries, phase, title, overall status, primary purpose \\
\textbf{Pubmed} & Publications associated to clinical trials \\
\textbf{Knowledge graph} & Parent-child relationships between diseases taken from UMLS, disease concepts (with the diseases' preferred term and synonyms), language. \\
\end{tabular}
\end{table}

Table \ref{tab:methods_data_used} shows the properties, from different data sources, that were used to engineer features. The properties, by themselves, do not measure how explainable a clinical trial is. Therefore, features were created using the conditions graph, the UMLS and pubmed databases to assess how much of a clinical trial the AI can explain. The engineered features are provided in Table \ref{tab:features}.

\begin{table}
\centering

\caption{Classification of features created for the local explainability based search engine.}
\label{tab:features}

\begin{adjustbox}{width=0.5\textwidth}
\begin{tabular}{|p{2cm}|p{4cm}|p{4cm}|p{2cm}|}
\hline
\textbf{Feature output type/ Query dependency} & \textbf{Query dependent} & \textbf{Query independent} & \textbf{Output}\\
\hline
\textbf{Binary} & query in title, preferred term in title & clinical stage present, stage is recruiting, overall status given & 0 or 1\\
\textbf{Numeric} &  query in summary, preferred term in summary, preferred term in summary, query in detailed description & number of publications & Between 0 and infinity\\
\hline
\end{tabular}
\end{adjustbox}
\end{table}

To facilitate the explainability-based calculations, features were assigned to various categories. Table \ref{tab:features} shows different classifications where the first category is based on the user's query: the feature is either query dependent or query independent. For example, the feature \textit{query in title} is a query dependent feature as the feature depends on a match between a query and the title of the clinical trial. So, if a clinical trial about breast cancer mentions the condition in the title, the feature receives a score of 1.  In contrast, query independent features do not depend on the query as regardless of what the query is, its score remains unchanged. For example, the feature \textit{number of publications} attributed to a clinical trial remains fixed, regardless of user's query. Second, Table \ref{tab:features} shows engineered features have two distinct outputs which are either binary or numeric. Binary features assess the presence of a feature in a study. Therefore, its output is either 0 or 1. On the other hand, numeric features count the occurrence of a feature. Thus, its outputs range between 0 and infinity.

%%%%%%%%%%%%%%%%%%%%%%%%%%%%%%%%%%%%%%% Feature Importance Identification
\subsection{Feature Importance Identification: A Crowdsourcing Approach} \label{crowdsourcing_approach}

%Make the statistical analysis as a methodology for weighting of the features and comparing them! 

We created a statistical approach to determine the weights of our features, and conducted a crowdsourcing task on Amazon Sagemaker as an alternative method to collect data on feature importance. Compliance regulations prohibits pharmaceutical companies to retain data on its users, especially when these relate to drugs \cite{Clinical58:online}. However, previous research has shown that crowdsourcees provided equal quality answers when conducting medical labeling tasks compared to domain experts \cite{dumitrache2013dr,dumitrache2017crowdsourcing}. Hence, in this experiment, 1116 responses were collected from participants to determine users' feature preferences and, thereupon, use these to order and explain results returned by the clinical search engine. 

Features' importance were measured using a cold start implicit strategy, where we asked participants to rate explainability sentences. The rating consisted of assessing sentences on a 5-point Likert-scale, from "Not convincing at all" to "Very convincing". Each sentence explained the prominence or availability of a feature mentioned in Table \ref{tab:features}. In addition, we changed the format of the sentences to implicitly measure which sentence format was most preferred to users in order to create user-friendly explanations. Therefore, when participants were asked to rate how convincing an explanation was to continue to read the clinical trial in further detail, we were implicitly measuring how important a certain feature was for our users.

We hypothesized that search features are not equally preferred by users. In addition, we hypothesized that the formulation of explanations were not equally preferred. We tested three dimensions of sentence formulation:  numeric  vs.  non-numeric  (using entities ‘3  times’ vs. ‘multiple  times’ in  an explanatory  sentences), action-oriented versus fact-driven formulations (‘retrieved’ versus ‘clearly mentioned’), and disease specific versus non-disease specific outputs (‘HIV’ versus ‘condition’).

\subsubsection{Results} \label{crowdsourcing_results}

\begin{table}[H]
\centering
\caption{Features' means and standard deviations.}\label{results_means-stdev}

\begin{tabular}{|l|p{1.5cm}|p{1.5cm}|}
\hline
\textbf{Features} & \textbf{Mean} & \textbf{Std dev}  \\
\hline
\textbf{Query in detailed description} &  3.69 & 0.82 \\
\textbf{Query in summary} &  3.53 & 0.91  \\
\textbf{Primary purpose availability} &  3.53 & 0.84  \\
\textbf{Number of publications} &  3.51 & 0.92  \\
\textbf{Stage availability} &  3.44 & 0.99  \\
\textbf{Query in title} &  3.15 & 0.93  \\
\textbf{Trial is recruiting} &  3.13 & 1  \\

\hline
\end{tabular}
\end{table}

\begin{table}[H]
%\centering

\caption{Results of feature labeling task. Note: the results in bold are statistically significant under the assumption p \textless 0.05.}\label{results_chi-square_results}
\begin{adjustbox}{width=0.5\textwidth}

\begin{tabular}{|l|l|l|l|l|l|l|l|}
\hline
\textbf{Features} & \textbf{Title} & \textbf{Summary} & \textbf{Description} & \textbf{Publications} & \textbf{Stage} & \textbf{Recruiting} & \textbf{Primary purpose}\\
\hline
\textbf{Title} & / & / & / & / & / & / & / \\
\textbf{Summary} & \textbf{\textit{0.007}} & / & / & / & / & / & / \\
\textbf{Description} & \textbf{\textit{0.00002}} & 0.51 & / & / & / & / & / \\
\textbf{Publications} & \textbf{\textit{0.02}} & 0.61 & 0.35 & / & / & / & / \\
\textbf{Stage} & \textbf{\textit{0.038}} & 0.67 & 0.15 & 0.88 & / & / & / \\
\textbf{Recruiting} & 0.81 & \textbf{\textit{0.006}} & \textbf{\textit{0.00001}} & \textbf{\textit{0.01}} & 0.054 & / & / \\
\textbf{Primary purpose} & \textbf{\textit{0.012}} & 0.82 & 0.42 & 0.82 & 0.43 & \textbf{\textit{0.006}} & /  \\

\hline
\end{tabular}
\end{adjustbox}
\end{table}

The results suggest that, in response to \textbf{feature importance}, partial ordering can be obtained via crowdsourcing tasks and statistical tests.  The results in Table \ref{results_means-stdev} illustrate that the feature with the highest mean score (3.69, on a 5-point Likert scale) was \textit{Query in detailed description}, whereas the two least convincing features were \textit{Query in title}, and \textit{Trial is recruiting} (3.15, and 3.13, respectively). We determined the weights of our features using $\chi^2$ tests. Table \ref{results_chi-square_results} provides the results of these chi-square tests where, for example, the features \textit{Query in title} and \textit{Query in summary} were not equally preferred (as the results reveal a p-value of 0.007). 

Data obtained in response to \textbf{feature importance} shows that there is at least a partial ordering that can be obtained via the crowdsourcing (based on statistical tests).   We determined the features' weight, using $\chi^2$ tests, based on statistical values. If two features were, for example, not statistically equally preferred, these two features would be attributed different weights. 

\begin{table} [H]
\centering
\caption{Experiment results for entities. Note: the results in bold are statistically significant under the assumption p \textless 0.05.}\label{results_entities}

\begin{adjustbox}{width=0.5\textwidth}

\begin{tabular}{|p{5cm}|p{1.5cm}|p{1.5cm}|p{1.8cm}|}
\hline
\textbf{Entity} & \textbf{$\overline{x}$(1)} & \textbf{$\overline{x}$(2)} & \textbf{P-value} \ \\
\hline
\textbf{(1) Non-numerical} 

\textbf{(2) Numerical} & 3.7 & 3.34 & \textbf{0.01}\\
\textbf{(1) Clearly mentioned} 

\textbf{(2) Retrieved}& 3.65 & 3.33 & \textbf{0.036} \\

\textbf{(1) Specify disease} 

\textbf{(2) Not specify disease} & 3.4 & 3.48 & 0.44\\

\hline
\end{tabular}
\end{adjustbox}
\end{table}

When it comes to results for the three \textbf{formulation} dimensions (Table \ref{results_entities}), when performing $\chi^2$ tests, we found that: 

\begin{itemize}
\item Users prefer explanations without non-numerical sentences (e.g. sentences mentioning that there are \textit{'multiple'} articles linked to the clinical trial vs. \textit{'2'}). 

\item Users prefer factual sentences (\textit{`clearly mentioned'}) compared to actions related to the search procedure (\textit{`retrieved'}).

\item No preferences were found between specifying the condition in a sentence  (e.g. the condition \textit{`HIV'} was mentioned in the title) versus (\textit{`the condition'}). 

\end{itemize}

%%%%%%%%%%%%%%%%%%%%%%%%%%%%%%%%%%%%%%% Explainability Score:
\subsection{Explainability Score: Ordering Retrieved Items} \label{explainability_scores}

In this section, we use the importance of the extracted features to compute the explainability score for each of the clinical trials and order them accordingly. For doing so, features are first assigned a weight, which are then used to calculate the explainability score, and ultimately the clinical trials are ordered based on these scores. 

Each clinical trial was attributed an explainability score based on its features availability or occurrences. Certain scores were fixed, while others depended on the user's query. The former are defined as query independent features, and the latter as query dependent features. As such, query dependent and query independent features were separately calculated (Table \ref{tab:features} reports which features belong to each category). 

Although query dependent and query independent scores were separately calculated, all explainability feature scores, shown as $e_f$, were calculated in the same manner: 

\[e_f = w * f_s\]

where the explainability score for each feature is determined by the weight $w$ (which depends on if the feature is binary or numeric, and if it is high or low importance), and the feature's score $f_s$. Binary  $f_s$ scores are identically determined for all binary features. If the feature is present; $f_s = 1$, and if the feature is unavailable; $f_s = 0$. However, numerical features $f_s$ scores are calculated based on each feature's occurrence. 

%\textbf{Should we discuss in more detail how the explainability score is obtained for query-dependent and query-independent features? You have it as comments}
As previously mentioned, the process to calculate the scores differ for query dependent and query independent features. For query dependent features, because all the terms related to one CUI refer to the same condition, all $e_f$ scores related to one CUI were grouped per clinical trial: 

\[E_{dtc} = \sum e_{f_{dtc}}  \] 

where the explainability scores $E$ for features belonging in the category query dependent $d$ were calculated by grouping features' score per CUI $c$ for each clinical trial $t$. On the other hand, features' scores belonging to the query independent category $i$ scores were calculated as: 

\[E_{it} = \sum e_{f_{it}}  \]
 
where all the $E_{i_t}$ scores are attributed to their respective studies, and linked to all the study's CUIs. Therefore, as long as the HCP queries a condition related to the clinical trial, the $E_{i_t}$ score attributed to that clinical trial will remain unchanged. Finally, this score will be summed per CUI with its $E_{d_{t_c}}$, which will give us our final explainability per clinical trial per CUI: 

\[E_{ct} = E_{it} +  E_{dtc} \]

where $E_{c_t}$ was used to order the clinical trials by conducting linear feature ranking. Explainability scores $E_{c_t}$ range between 0 and 1, where clinical trials with scores close to 1 reflect that the search engine can explain more about these clinical trials compared to clinical trials with a score close to 0. Hence, the clinical trials linked to a CUI (that is queried by the user) with the highest XAI scores for that CUI appear higher in the results' list. 

Therefore, the algorithm orders the clinical trials by their explainability scores $ef$. To do so, the algorithm takes as input the HCP's query condition. The algorithm will then search for the query term in the database, and identify the CUI associated to the condition. Secondly, the algorithm filters all clinical trials to keep studies related to that CUI, therefore providing a list of all articles related to the condition the HCP queried. Lastly, the list will be ordered based on explainability scores $ef$, where the highest explainable scores will receive the highest ordering position, and the lowest explainability scores will receive the lowest position. 

%%%%%%%%%%%%%%%%%%%%%%%%%%%%%%%%%%%%%%% Recommender Explanations
\subsection{Retrieval Explanations}
Having extracted the features and computed their importance, this section is dedicated on how to explain the retrieved items in response to a query from the user. The explanations must be in a way that the HCP can readily understand them. For doing so, we develop a a template list of sentences, as shown in Table \ref{tab:template sentences}. These sentences are simple, user-friendly, hierarchically structured, short and straightforward, as well as they can explain the source of information, and cover why a result was returned. 

\begin{table} [H]
\centering
\caption{Template sentences created for the explainability based search engine.}\label{tab:template sentences}
\begin{adjustbox}{width=0.5\textwidth}

\begin{tabular}{|l|l|}
\hline
\textbf{Feature} & \textbf{Template sentence}   \\
\hline
\textbf{Query in title} & The condition is mentioned in the title \\
\textbf{Preferred term in title} & The preferred term of the condition is mentioned in the title  \\
\textbf{Query in summary} &  The condition is mentioned in the summary  \\
\textbf{Preferred term in summary} &  The preferred term of the condition is mentioned in the summary  \\
\textbf{Query in detailed description} & The condition is mentioned in the detailed description  \\
\textbf{Preferred term in detailed description} & The preferred term of the condition is mentioned multiple times in the description \\
\textbf{Number of publications} & The clinical trial has multiple publications \\
\textbf{Stage availability} & The clinical trial's stage is clearly mentioned  \\
\textbf{Overall status availability} & The clinical trial's status is clearly mentioned \\
\textbf{Trial is recruiting} & The clinical trial's status is recruiting \\

\hline
\end{tabular}
\end{adjustbox}

\end{table}

%It is important to note that the sentences are generated based on the features identified by the engine. As presented in the process pipeline 2 in Figure \ref{features_scores_pipeline}, these are not related to explainability feature scores, hence they are not dependent on the weights, but are only related to the features themselves.

Sentences were created in the following manner: 
\begin{enumerate}
    \item \textit{A maximum of three sentences at a time are displayed.} We assume that, given the limited amount of time HCPs spend on the search engine, a maximum of three sentences will be enough for the HCP to read. 
    \item \textit{Sentences are only displayed if certain conditions are met.} Given the limited time this research has, thresholds are determined based on intuitive knowledge. This allows users to only see relevant explanations. 
    \item \textit{Sentences are ordered by feature preference.} The ordering at which sentences are displayed rely on the results of the experiment described in sub-section \ref{crowdsourcing_results}. 
    \item \textit{The sentences are kept simple.} To understand which formulation of sentences users prefer, we researched entity preferences as described in sub-section \ref{crowdsourcing_approach}. 
\end{enumerate}
%%%%%%%%%%%%%%%%%%%%%%%%%%%%%%%%%%%%%%%%%%%%%%%%%%%%%%%%%%%
%%%% Experiments
%%%%%%%%%%%%%%%%%%%%%%%%%%%%%%%%%%%%%%%%%%%%%%%%%%%%%%%%%%%
\section{MODEL EVALUATION}\label{sec:experiments}

We evaluated our model by comparing it to other simulated clinical search engines based on users' trust, search experience, and result ordering satisfaction. Our hypotheses were that all search engines were equally preferred in all three dimensions. We, therefore, simulated 5 different search engines with different city names, where each engine queried either \textit{lyme disease}, \textit{breast cancer}, or \textit{HIV}:
\begin{enumerate}
    \item\textbf{Amsterdam}: Search engine \textit{with} ordered results and \textit{with} explainable sentences 
    \item \textbf{Berlin}: Search engine \textit{with} ordered results and \textit{without} explainable sentences 
    \item \textbf{Copenhagen}: Search engine \textit{without} ordered results and \textit{with} explainable sentences 
    \item \textbf{Dublin}: Search engine \textit{without} ordered results and \textit{without} explainable sentences
    \item \textbf{Edinburgh}: Search engine with titles ordered by alphabetical order
\end{enumerate}
The engines used data from myTomorrows\footnote{https://search.mytomorrows.com/public} in order to create scenarios as realistic as possible. The different query concepts were queried in each search engine, for which the top 10 results were extracted and put into the simulated environments to imitate the first page of a search engine showing 10 results at a time.

\subsection{Experiment setup} \label{experiment_setup}
Participants were recruited using different social media platforms such as Facebook, Linkedin, or recruited in the company itself. Participants received a link to a questionnaire focusing on one of the query concepts (either \textit{lyme disease}, \textit{HIV}, or \textit{invasive breast cancer}). In each questionnaire, participants were shown one by one the different simulated search engines related to the query concept. The different search engines were shown in a random order. Additionally, participants were asked to: 
\begin{itemize}
    \item Assess if they \textbf{trusted }the search engine. 
    \begin{itemize}
        \item Question asked: \textit{When looking at the search engine, how much do you trust the search engine?}
        \item Possible answers: 
        \begin{enumerate}
            \item \textit{I trust this search engine very much}
            \item\textit{ I trust this search engine}
            \item \textit{My trust is neutral}
            \item \textit{I do not trust this search engine}
            \item \textit{I do not trust this search engine at all}
        \end{enumerate}
    \end{itemize}
    \item Assess if they were satisfied with the \textbf{ordering} of the search engines' results 
    \begin{itemize}
        \item Question asked: \textit{When looking at the search engine, are you satisfied with the ordering of clinical trials? }
        \item Possible answers: 
        \begin{enumerate}
            \item \textit{I am very satisfied with the ordering}
            \item\textit{ I am satisfied with the ordering}
            \item\textit{ I feel neutral }
            \item \textit{I am not satisfied with the ordering}
            \item \textit{I am highly not satisfied with the ordering}
        \end{enumerate}
    \end{itemize} 
    \item Asses their \textbf{search experience} while using the search engine
    \begin{itemize}
        \item Question asked: \textit{What is your search experience when using the search engine?}
        \item Possible answers: 
        \begin{enumerate}
            \item \textit{I have a great search experience}
            \item \textit{I have a good search experience }
            \item \textit{My search experience is neutral} 
            \item \textit{My search experience is not good }
            \item \textit{My search experience is not good at all}
        \end{enumerate}
    \end{itemize}
\end{itemize}

In the end of the questionnaire, participants were asked to order the different search engines by: 
\begin{itemize}
    \item \textit{Trust}: users had to order the search engines from most trustworthy to least trustworthy.
    \item \textit{Result ordering satisfaction}: users were asked to order of search engines they preferred from highest result ordering satisfaction to lowest result ordering satisfaction.
    \item \textit{Search experience}: users were asked to order the search engines from best search experience to least favourite search experience.
\end{itemize}

\begin{figure} [H]
\centering
\includegraphics[width=0.5\textwidth]{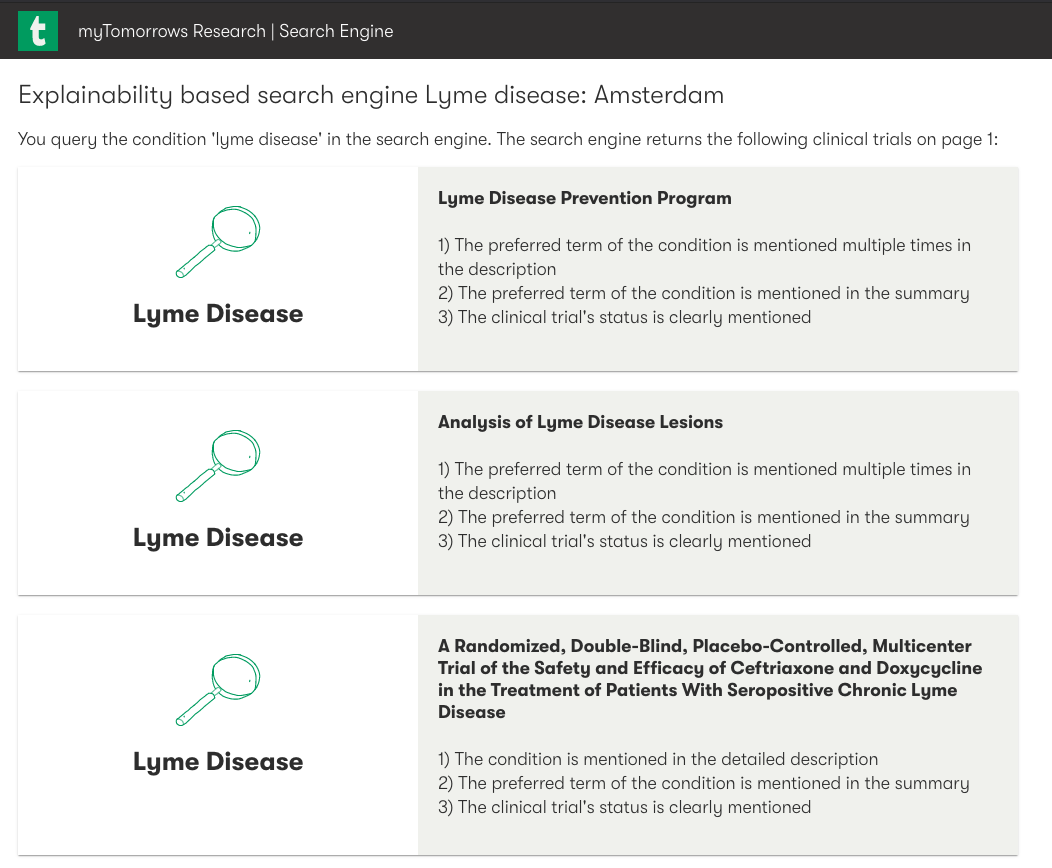}
\caption{Example of a simulated search engine.} \label{simulated_search_engine}
\end{figure}

An example of one of the simulated search engines is shown in Figure \ref{simulated_search_engine}. 

In total, we created 9 questionnaires which were randomly allocated to 55 participants, among which 34 completed the experiment. 
%55 questionnaires were sent out to participants where 34 completed the experiment, among which *** were HCPs, and *** were non-HCPs. 

\subsection{Results} 

\begin{table}[H]
\centering
\caption{$\chi^2$ results of the comparison of all search engines to each other.}\label{results_table_chi-square}
\begin{tabular}{|c|c|c|c|}
\hline
\textbf{} & \textbf{All participants} & \textbf{HCPs} & \textbf{Non-HCPs}\\
\hline
\textbf{Trust} & 0.29 & 0.44 & 0.62 \\
\textbf{Search experience} & \textbf{\textit{0.04$*$}} & 0.09 & 0.37 \\
\textbf{Ordering} & 0.31 & 0.40 & 0.48 \\
\hline
\end{tabular}

\small Note: the results in \textbf{\textit{bold italic with a *}} are statistically significant under the assumption p \textless 0.05.
\small By \textit{all participants,} we mean the combination of results of HCPs and non-HCPs. 
\end{table}

 In the experiment, we asked participants to evaluate the different search engines one by one and report their experience. We conducted the $\chi^2$ test to test our hypotheses that all search engines are equally preferred in all three dimensions. Table \ref{results_table_chi-square} reports the $\chi^2$ test results, and shows that when combining the responses of HCPs and non-HCPs, the null hypothesis that all search engines have equal search experience is rejected as the test returned a p-value of 0.04. This suggests that users have different search experiences when, distinctively, facing the search engines. The three following subsections investigate how all participants (HCPs and non-HCPs), HCPs alone, and non-HCPs alone, evaluate the different search engines.

\begin{figure}[htp]
\centering
\includegraphics[width=0.5\textwidth]{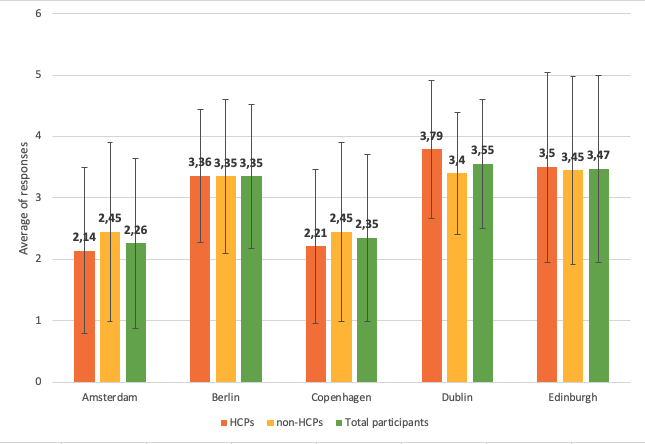}
\caption{Results: order the search engines from \textit{most \textbf{trusted}} to \textit{least \textbf{trusted}}.} \label{figure_results_order_trust}
\small Note: Results close to 1 indicate the most preferred search engines. On the contrary, results close to 5 are the least preferred search engines. 
\end{figure}

\begin{figure} [htp]
\centering
\includegraphics[width=0.5\textwidth]{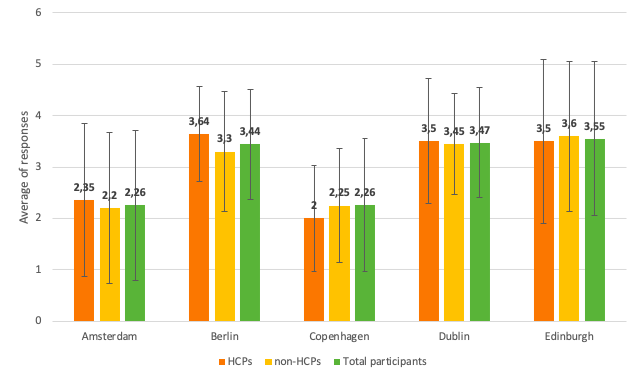}
\caption{Results of ordering task: order the search engines from \textbf{\textit{best search experience}} to  \textbf{\textit{worst search experience}}.} \label{figure_results_order_searchexperience}
\small Note: Results close to 1 indicate the most preferred search engines. On the contrary, results close to 5 are the least preferred search engines. 
\end{figure}

 The results of the task asking participants to order the search engines from most trusted to least trusted, and best to worst search experience, suggest that users, both HCPs and non-HCPs, reported more trust and better search experience while using the search engines using explainability sentences (Amsterdam and Copenhagen) compared to search engines not explaining its results (Berlin, Dublin and Edinburgh). The results displayed in Figure \ref{figure_results_order_trust} and Figure \ref{figure_results_order_searchexperience} demonstrate these preferences. 
 
The results of the task asking participants to order the search engines from best result ordering satisfaction to worst ordering satisfaction are reported in in Figure \ref{figure_results_order_ordering}. Similar to the results for search experience, search engines including explanations of results improve ordering satisfaction (Amsterdam and Copenhagen). In addition, HCPs ranked last the search engine with explainability based ordering without explanations (Berlin). This further demonstrates that failing to explain how the results' ordering works leads to reduced result ordering satisfaction. 

%%%%%%%%%%%%%%%%%%%%%%%%%%%%%%%%%%%%%%%%%%%%%%%%%%%%%%%%%%%
%%%% Conclusion
%%%%%%%%%%%%%%%%%%%%%%%%%%%%%%%%%%%%%%%%%%%%%%%%%%%%%%%%%%%
\section{Discussion}\label{sec:conclusion}

When asked to report the preferred order of search engines, participants consistently preferred in all three dimensions the search engines Amsterdam and Copenhagen. The two search engines include explainability based sentences and ordering, and explainability based sentences, respectively. However, we noticed that search engine Berlin scored low, even last in the dimension of ordering satisfaction with HCPs. This suggests that explainability based ordering is preferred when explained with user-friendly sentences. This is in line with research in \cite{pu2006trust} as authors demonstrated that explainability overall increases trust. A reason is that without explainability sentences, users understand less the logic behind the model, and therefore attribute lower scores to the search engine Berlin. This reasoning is further emphasized by the significant difference in ordering preference between Dublin and Edinburgh for non-HCPs, where these prefer Edinburgh given that the logic of the search engine is  straightforward, which can increase user satisfaction. 

HCPs' results on user experience, trust and ordering satisfaction could be influenced by their expertise. When assessing the different dimensions of the search engines, HCPs could be looking for familiarity and, therefore, search for clinical trials that are within their field of experience. For example, radiologists would pursue clinical trials related to radiology, etc. In addition, a common problem in the medical domain are discrepancies in patient diagnosis between HCPs. Research has shown that subjective preferences influence a diagnosis' outcome \cite{gierada2008lung}, which could explain why HCPs have high variety with their responses when diagnosing a patient. This follows that subjective preferences could have influenced the results of the evaluation of our explainability-based search engine.  

Although the model is scalable and generalizable, the features created for this use case are not transferable to other search engines.  Features need to adapt to other models' use cases as most features created in this research are specific to clinical trials. For exmple, a search engine returning a list of travel destinations would not benefit from the feature \textit{'the clinical trial is recruiting'}. Transferring the model as-it-is to another set of data would, therefore, require adapting the method to the different use-case. In addition, developers would need to collect data on feature preferences for their use-case.

\section{Future work}

This research aimed to measure the influence of explainability based search engines on users’ trust, search experience and result ordering satisfaction. Two experiments were conducted, where the first experiment was created to order explainability based features based on importance. The results were translated to weights for features to order the results returned by the search engine. The second experiment evaluated the explainability based search engine by measuring users’ experience with the engine. Overall, the results suggest that search engines with explanations are more trusted, provide greater user experience, and increased ordering satisfaction, compared to search engines without explanations. In addition, users are satisfied with explainability based ordering of results if these have explanations, where not explaining the ordering of results decreases users’ trust, search experience and ordering satisfaction. Thus, the results urge developers to explain search engine results. 

Recommended future work is to investigate if there is a need for different explanations for HCPs compared to non-HCPs. HCPs could require more detailed explanations such as additional information on sources such as UMLS and Pubmed. On the contrary, non-HCPs would rather have less medical terms in order to interpret the explanations. Therefore, investigate the different needs of explanations depending on the users' background would benefit explainable research.

Although our model provides explainability sentences, these are not personalized to the profile of the HCP. In order to make it more personal, results could be ordered based on the HCP’s profile and preferences. To achieve this, future work should collect data on user profiles, and use machine learning to identify users' personal preferences. Moreover, this could be combined with knowledge graphs to create a relationship between clinical trials and users' profiles as shown in \cite{catherine2017explainable}, where users were provided personal explanations using a knowledge graph based on item reviews and user profile.

\bibliographystyle{IEEEtran}
\bibliography{reference}
\end{document}